\documentclass[12pt]{article}
\usepackage{amssymb,graphicx,epsfig} 
\def\baselinestretch{1.2}

\begin{document}
\begin{titlepage}
\begin{flushright}
TIFR/TH/10-16
\end{flushright}

\bigskip\bigskip

\begin{center}
{\Large {\bf
{Boosted Top Quark Signals for Heavy Vector Boson   \\ [3mm]
Excitations in a Universal Extra Dimension Model }}} \\[5mm]
\bigskip
{\sf Biplob Bhattacherjee} $^{a,}$\footnote{biplob@theory.tifr.res.in}, 
{\sf Manoranjan Guchait} $^{b,}$\footnote{guchait@tifr.res.in}, 
{\sf Sreerup Raychaudhuri} $^{a,}$\footnote{sreerup@theory.tifr.res.in} 
and 
{\sf K. Sridhar} $^{a,}$\footnote{sridhar@theory.tifr.res.in} \\ [4mm]

\rm $^a$ Department of Theoretical Physics, \\
Tata Institute of Fundamental Research, \\
1 Homi Bhabha Road, Mumbai 400 005, India. \\

\bigskip

\rm $^b$ Department of High Energy Physics, \\
Tata Institute of Fundamental Research, \\
1 Homi Bhabha Road, Mumbai 400 005, India. \\

\normalsize
\vskip 30pt

{\large\bf ABSTRACT}
\end{center}

\begin{quotation} \noindent
{\small In view of the fact that the $n = 1$ Kaluza-Klein (KK) modes in 
a model with a Universal Extra Dimension (UED), could mimic 
supersymmetry signatures at the LHC, it is necessary to look for the $n 
= 2$ KK modes, which have no analogues in supersymmetry. We discuss the 
possibility of searching for heavy $n = 2$ vector boson resonances -- 
especially the $g_2$ -- through their decays to a highly-boosted top 
quark-antiquark pair using recently-developed top-jet tagging techniques 
in the hadronic channel. It is shown that $t\bar{t}$ signals from the $n 
= 2$ gluon resonance are as efficient a discovery mode at the LHC as 
dilepton channels from the $\gamma_2$ and $Z_2$ resonances. }
\end{quotation}

\bigskip 

\begin{center} 
PACS numbers: {\tt 14.80.Rt, 14.65.Ha} 
\end{center}

\vskip 20pt
\begin{center}\today\end{center}

\vfill

\end{titlepage}
\newpage
\setcounter{page}{1}

\noindent Extra spatial dimensions, introduced rather tentatively in 
physics during the period 1914 -- 1926 \cite{Nordstrom1914, Kaluza1921, 
Klein1926}, had been more or less relegated to the category of arcana 
till the late 1990's, when they were re-introduced by Arkani-Hamed, 
Dimopoulos and Dvali \cite{ADD1998, RS1999} as a possible solution to 
the insidious hierarchy problem that plagues the Standard Model (SM) and 
many of its extensions. The subsequent decade saw an outburst of 
creativity in the field of model building using this concept. Much of 
that effort was driven by the hope that having extra dimensions of 
different shapes and sizes, and locating the SM particles in different 
subspaces of these, could provide the long missing solutions to many of 
the puzzles inherent in the standard electroweak model. Today, a dozen 
years after the original proposal \cite{ADD1998}, some of the early 
euphoria has worn off. A sober appraisal will show that now there exist 
many offshoots of the original proposal, each having its own strengths 
as well as its own drawbacks. One of the most attractive of these 
suggestions is the model with a single Universal Extra Dimension (UED-5) 
\cite{Appelquist2001}, which can be said to provide a viable new physics 
alternative with a {\it minimum} of new assumptions. As this paper is 
devoted to considering some signatures of this UED-5 model, a short 
introduction to its main features seems appropriate.

\bigskip\noindent The UED-5 model is rather close to the primitive model 
of Kaluza and Klein\cite{Appelquist1989}, in that it envisages a single 
extra compact dimension $x^4$ of space, and allows all the fields of the 
SM to propagate in this extra dimension, as well as in the canonical 
four dimensions of Minkowski space. It differs, however, from standard 
Kaluza-Klein (KK) theory in two crucial aspects:
\begin{enumerate}

\item Though the metric tensor contains off-diagonal elements which 
constitute a four-vector, no attempt is made to identify this with any 
of the gauge bosons of the SM. This sidesteps an important stumbling 
block of the KK theory and enables the extra dimension to be much larger 
than the Planck length. Interactions due to these off-diagonal elements 
will be completely negligible for elementary particle physics at 
laboratory energies and need not be considered further.

\item The extra fifth dimension $x^4$ is compactified, with the 
topology, not of a circle $\mathbb{S}^{(1)}$, as in the KK theory, but 
of a circle folded about one of its diameters, i.e. an `orbifold' 
$\mathbb{S}^{(1)}/\mathbb{Z}_2$. It is well known \cite{Witten1981} that 
a fifth dimension in the form of a manifold -- in this case a simple 
circle $\mathbb{S}^{(1)}$ -- after compactification, cannot give rise to 
chiral fermions in four dimensions, whereas a 
$\mathbb{S}^{(1)}/\mathbb{Z}_2$ orbifold can. Thus, the UED model can 
accommodate the chiral quarks and leptons of the Standard electroweak 
theory, which the primitive KK theory could not.

\end{enumerate}

\noindent These two changes are enough to permit the construction of a 
model having heavy KK excitations of every SM particle. After 
compactification, at tree-level, the particle masses $M_n$ of the $n$-th 
order KK excitation are given by
\begin{equation}
M_n^2 = M_0^2 + n^2 R^{-2}
\end{equation}
where $M_0$ is the mass in five dimensions and $R$ is the radius of 
compactification of the extra dimension. These KK masses are, of course, 
determined by the momentum in the fifth direction, which is necessarily 
discretised as $p_4^{(n)} = n R^{-1}$ by the periodic boundary 
condition. The integer $n$ is referred to as the \textit{KK number}. The 
$n = 0$ modes (i.e. those which never go into the fifth dimension) are 
identified with the SM particles.

\bigskip\noindent The presence of large numbers of KK excitations leads 
to dramatic consequences when we consider the running of the SM gauge 
coupling constants $g_1$, $g_2$ and $g_3$. Since the beta functions 
change every time a KK threshold $R^{-1}, 2R^{-1}, 3R^{-1}, \dots $ is 
crossed, the running over large energy ranges resembles a power-law 
behaviour rather than the usual logarithmic running \cite{Dienes1998}. 
It can be shown that this leads to a meeting (within experimental error) 
of the three coupling constants at a scale of $\Lambda \simeq 20 
R^{-1}$, i.e. after crossing about 20 KK thresholds. Accordingly, the 
minimal UED-5 model should be cut off at the scale $20 R^{-1}$, or 
earlier, where one would expect a unified gauge group to take over. Even 
if we do not believe in grand unification and put down the triple 
meeting of the coupling constants to be a coincidence, the theory must 
definitely be cut off around 40 thresholds, since at such energies the 
U(1) coupling $g_1$ develops a Landau pole. Note that with a typical 
experimentally allowed value of $R^{-1} \approx 500$~GeV, the 
unification point is $\Lambda \approx 10$~TeV, which is rather beyond 
the kinematic range of the CERN LHC, but far below the traditional grand 
unification scale. Though serious model building has not really been 
done in this context, it is obvious that, with such a low unification 
scale, any grand unified gauge group must include some discrete symmetry 
to prevent ultra-fast proton decay. However, in the present work we are 
concerned with the theory well below the cutoff $\Lambda$, which 
includes lepton and baryon number conservation exactly as in the SM, and 
hence, such issues are not of primary concern.

\bigskip\noindent Thus, there are two unknown parameters in the theory. 
One is $\Lambda R$, for which the most reasonable value is around 20, 
but which can be taken as low as 5 (i.e. just beyond the LHC accessible 
range\footnote{This reflects the most conservative view that one should 
not extrapolate the theory beyond the kinematically accessible 
experimental limit.}) and as high as 40 (the Landau pole in $g_1$). The 
other free parameter is the size parameter $R^{-1}$, which controls the 
masses of the KK excitations.  Experimental constraints arising from the 
radiative decay $B \to X_s \gamma$ tell us \cite{B2Xg2007} that the 
value of $R^{-1}$ must definitely be above 300~GeV, and is likely to be 
above 600~GeV if the Higgs boson turns out to be as light as is hinted 
at by electroweak precision data \cite{lightHiggs}. There is no {\it 
experimental} upper bound, of course, but if $R^{-1} > 1400$~GeV, then 
the UED-5 model is unable, at the 90\% confidence level, to explain the 
dark matter density extracted from cosmic microwave background data 
\cite{LKP2007}. Taking all this into account, our choice of range is
$$
400~{\rm GeV} < R^{-1} < 1400~{\rm GeV} \qquad\qquad\qquad 5 < \Lambda R < 20
$$
which stretches the experimental and phenomenological limits within 
reason, but not to absolute extremes.

\bigskip\noindent The $\mathbb{S}^{(1)}/\mathbb{Z}_2$ orbifold possesses 
\textit{local} translation invariance, as a result of which the momentum 
in the fifth direction $x^4$ is conserved at the tree level. Noting that 
$p_4^{(i)} = n_i R^{-1}$ for the $i$-th incoming particle entering into 
a reaction, momentum conservation $\sum_i p_4^{(i)} = 0$ also implies 
the conservation of KK number, i.e. $\sum_i n_i = 0$. However, the 
translation invariance is clearly broken \textit{globally} at the 
boundaries of the orbifold $\mathbb{S}^{(1)}/\mathbb{Z}_2$, i.e. at the 
two ends of the diameter about which folding has taken place. Thus, 
quantum corrections can break the translation invariance 
\cite{Cheng2002} if they involve virtual states which wind around the 
extra dimension, and are, therefore, sensitive to the size of the 
compact dimension. Moreover, such propagators receive contributions from 
the boundary conditions at the so-called orbifold fixed points. This has 
two immediate consequences at the one-loop level. The first is to 
provide additional degeneracy-lifting contributions\footnote{Though 
these radiative corrections do increase the mass splitting between $n = 
1$ modes of different fields, the splitting ($< 30$\%) is not very large 
since it is, after all, a perturbative effect.} to the masses $M_n$, 
apart from the $M_0$. These have been calculated by Cheng \textit{et al} 
\cite{Cheng2002} and the corresponding spectrum has recently been 
automated in a {\sc CalcHEP} framework by Datta \textit{et al} 
\cite{Datta2010}. It is important to note that while the overall 
splitting more or less scales as $R^{-1}$, it also grows logarithmically 
with $\Lambda R$ -- which is not unexpected since all the loop momenta 
have to be cut off at $\Lambda$.

\bigskip\noindent The second -- and more far-reaching -- consequence of 
the breaking of translation invariance is that it is now possible to 
have non-conservation of momentum in the fifth direction, which implies 
non-conservation of KK number, i.e. it is possible to have $\sum_i n_i 
\neq 0$. Nevertheless, there still exists a $Z_2$ symmetry corresponding 
to interchange of the two ends of the folded circle, and this enforces 
conservation of a multiplicative quantum number $\eta = (-)^n$, which we 
call \textit{KK parity}. All the SM particles have $n = 0$ and hence 
possess even KK parity $\eta = +1$; all the $n = 1$ excitations have odd 
KK parity $\eta = -1$; all the $n =2$ modes again have even KK parity 
$\eta = +1$, and so on. If we focus on the $n = 1$ states, we note that 
they can neither be produced singly from SM particles nor decay 
individually into SM particles as that would led to non-conservation of 
KK parity. This is completely analogous to the way in which 
supersymmetric particle production and decay is controlled by the 
conservation of $R$ parity.

\bigskip\noindent The analogy with supersymmetry goes further, indeed 
--- for conservation of KK parity implies that the lightest of the $n = 
1$ states, the Lightest KK Particle (LKP), must be absolutely stable, 
just as the Lightest Supersymmetric Particle (LSP) is. Like the LSP, the 
LKP is also an excellent candidate for the dark matter component of the 
Universe \cite{LKP2007} --- in fact, the prediction of such a candidate 
is a strong \textit{\'a posteriori } motivation for the UED-5 model. 
Further, once we accept the fact that the LKP will be stable, the 
conservation of KK parity also ensures that it will interact only weakly 
with matter, by exchanging heavy $n = 1$ excitations, in the same way as 
the LSP exchanges only heavy supersymmetric particles, and neutrinos 
exchange heavy $W$ and $Z$ bosons. Because of these weak interactions 
with matter, the LKP will escape detection in terrestrial experiments 
and its appearance can only be inferred from missing transverse energy 
(MET) and momentum.

\bigskip\noindent At collider experiments, therefore, the UED-5 model 
bears a strikingly close resemblance to supersymmetry, for in both cases 
the SM particles have heavy undiscovered partners, of which the lightest 
will escape detection and be observable only as MET. This has lead to 
the UED-5 model being dubbed `bosonic supersymmetry' \cite{Matchev2002}. 
Most of the production and decay processes which occur in supersymmetry 
have their counterparts in the UED-5 model, as a result of which it is 
difficult, considering collider signals alone, to distinguish between 
supersymmetry and the UED-5 model. Some attempts to do this may be found 
in Ref.~\cite{Biplob2009}. These methods work well enough for a minimal 
SUGRA-inspired mass spectrum, but if we take the most general 
supersymmetric mass spectrum, it will be practically impossible to tell 
the difference between the two models using kinematic variables alone. 
The fact that the heavy partners in a UED-5 model carry the same spin 
whereas in a SUSY framework they have spin differing by a half unit, has 
also been proposed as a discriminator between the two models 
\cite{Reinartz2008}. However, the biggest difference between UED-5 and 
supersymmetric models lies in the fact that the UED-5 model predicts a 
whole tower of KK partners of each SM particle, whereas in any $N=1$ 
supersymmetric model, there is just one set of supersymmetric partners. 
Thus, additional discovery of one or more of the $n = 2$ KK modes of SM 
particles would be a `smoking gun' signal of the UED-5 model. As the $n 
= 2$ modes (for example) have KK parity $\eta = (-)^2 = +1$, they may be 
produced as resonances in the collision of SM particles. Moreover, 
because of the mass relation $M_2 \approx 2 R^{-1} \approx 2 M_1$, the 
energy required to produce a {\it pair} of $n = 1$ KK modes is roughly 
the same as that required to excite a {\it single} $n = 2$ KK mode 
resonance. This means that if we can produce a pair of $n = 1$ KK modes 
and see their cascade decays to the LKP -- which is the signal that 
mimics the signals for supersymmetry -- then, kinematically, we should 
also be able to produce $n = 2$ resonances. The combination of a 
supersymmetry-like signal with the existence of such a resonance would 
be a very strong signal, indeed, for a UED-5, and, therefore, it is 
necessary to consider the $n = 2$ KK mode resonances at the LHC in all 
seriousness.

\bigskip\noindent In order to produce $n = 2$ KK modes singly, we 
require to use their coupling to a pair of SM particles, i.e. to $n = 0$ 
modes. Such a vertex, which violates KK number but not KK parity, cannot 
be obtained at the tree level (i.e. with {\it local} operators) in the 
UED-5 model. However, as mentioned above such vertices are generated by 
one-loop diagrams of the form shown in Figure~\ref{fig:OneLoop}, which 
have $n = 1$ modes running in the loop and where every vertex is KK 
number-conserving, but one (or three) of the propagators violate KK 
number by crossing an orbifold boundary somewhere. At these boundaries 
the reflection symmetry in the fifth dimension ensures that the momentum 
$p_4$ -- and hence the KK number -- flips sign. The relevant vertices at 
the LHC would be of the form $q$--$\bar{q}$--$V_2$, where $V = \gamma, 
W, Z$ or $g$. These are readily available in the 
literature~\cite{Datta2010} and could lead to direct $s$-channel 
processes of the form $q\bar{q} \to V_2 \to f\bar{f}$, where $f$ is 
either a quark or a lepton. A similar class of vertex can be obtained by 
replacing the quarks in the above diagrams by leptons. However, for the 
purposes of this work, we neglect $g$--$g$--$g_2$ vertices, since these 
are removed, to leading order\footnote{There could be, in principle, 
higher dimensional operators for this vertex, which would be suppressed 
by a suitable high energy scale. The investigation of such effects, 
though interesting in itself, is postponed to a future work.}, by a 
re-diagonalisation procedure which is required so that the massless 
gluon states do not contain an admixture of massive $g_2$ states 
\cite{Ponton2005}. \\

\begin{figure}[htb]
\setcounter{figure}{0}
\centerline{ \epsfxsize= 3.8 in \epsfysize= 4.2 in \epsfbox{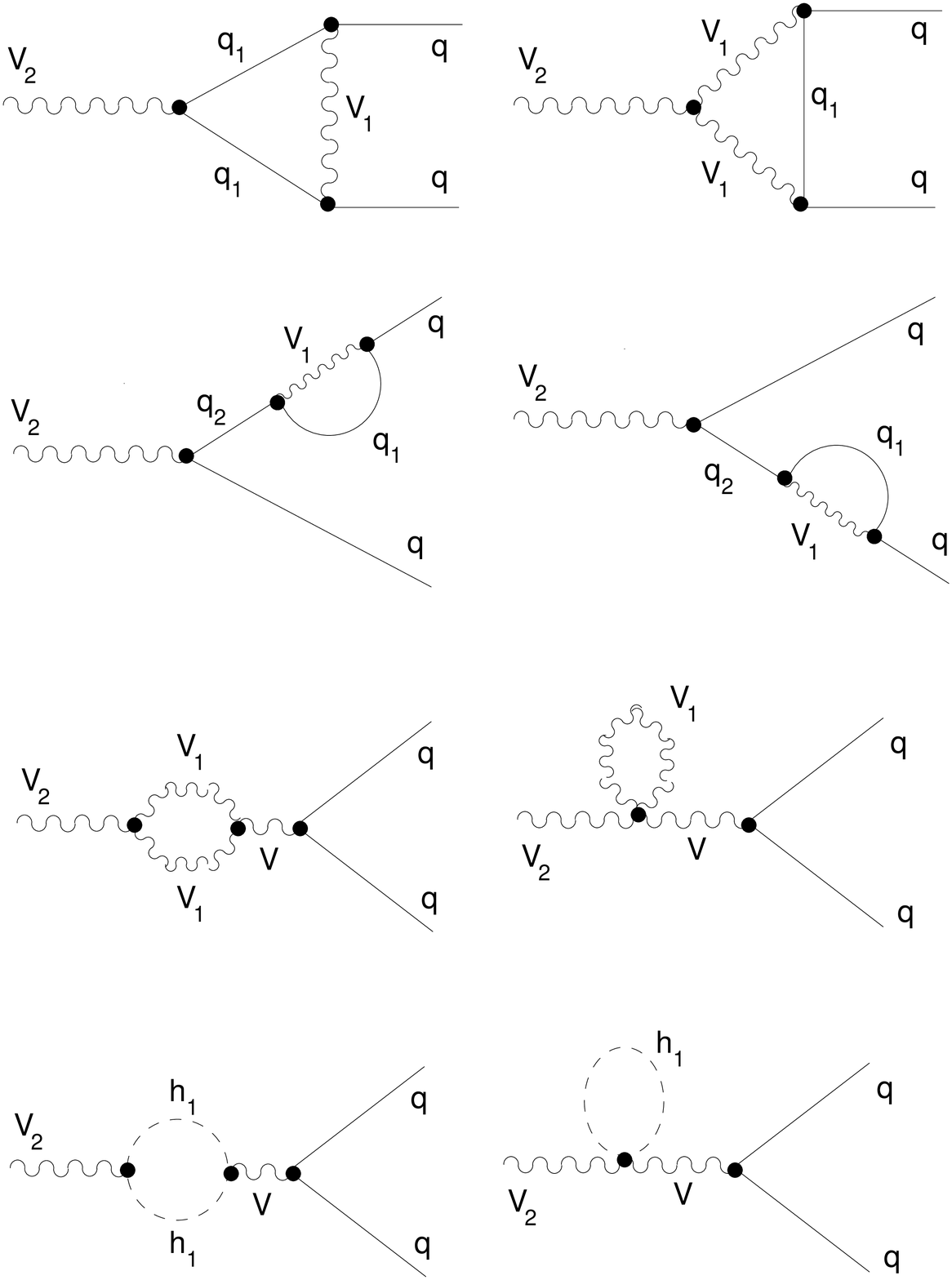} }
\vskip -10pt
\def\baselinestretch{0.8}
\caption{{\footnotesize\it One-loop diagrams giving rise to vertices of 
the form $q$--$\bar{q}$--$V_2$ in the UED-5 model. Here $q$ stands for 
any quark, and $V$ stands for any vector boson.  If $V_2 = g_2$, the 
diagrams on the fourth row are absent.}}
\def\baselinestretch{1.2}
\label{fig:OneLoop}
\end{figure}
\vskip -5pt

\noindent Granted the possibility of having vertices of the form 
$q$--$\bar{q}$--$V_2$, it follows that a $V_2$ boson can be produced 
on-shell at the LHC, kinematics permitting, through any of the following 
processes:
$$
u + \bar{u} \to \gamma_2, \ Z_2^0, \ g_2 
\qquad
d + \bar{d} \to \gamma_2, \ Z_2^0, \ g_2
\qquad
u + \bar{d} \to W_2^+
\qquad
\bar{u} + d \to W_2^-
$$
where the parton density function (PDF) of $u$ is understood to include 
the flux of $c$ while the PDF of $d$ includes those of both $s$ and $b$. 
Now, the same vertices, together with their leptonic counterparts, can 
be responsible for the decay processes
$$
\begin{array}{rcccccc}
\gamma_2, \ Z_2^0, \ g_2 & \to & \nu + \bar{\nu} \ , &  \ell^+ + \ell^- \ , & 
 q + \bar{q} \ , &  b + \bar{b} \ , &  t + \bar{t}  \\
W_2^+ & \to & \nu + \ell^+ \ , & & q + \bar{q}' \ ,& t +  \bar{b} \ , &  \\
W_2^- & \to & \bar{\nu} + \ell^- \ , & & q' + \bar{q} \ ,& & b +  \bar{t} \ ,   
\end{array}	
$$
where $\nu, \ell$ denote neutrinos and charged leptons, respectively, of 
any of the three generations, and $q, q'$ denote quarks of the first two 
generations. Of the final states which can be easily tagged at the LHC, 
the dilepton signals have already been studied in 
Refs.~\cite{Battaglia2005} and \cite{Datta2005}, while the dijet signals 
arising from light quarks will have an overwhelmingly large QCD 
background. At LHC energies, traditional $b$-tagging techniques, which 
rely on the reconstruction of displaced vertices, do not work well when 
the parent $b$-quark (or antiquark) is highly boosted and the decay 
products collimated into a narrow cone, especially when the $b$-jet has 
$p_T > 300$~GeV \cite{b-tagging}. However, when it comes to $t\bar{t}$ 
final states, it may be possible to detect resonances in them, provided 
we can efficiently reconstruct these massive quarks from their decay 
fragments. Tagging heavy quarks through jet sub-structure is a technique 
which has been recently been put to good use in new physics studies at 
the LHC \cite{t-tagging}, and, for top quarks, can be implemented easily 
enough using the software {\sc FastJet} \cite{FastJet}. Thus, in this 
work, we study $\gamma_2, \ Z_2^0$ and $g_2$ resonances in the process 
$p + p \to t + \bar{t}$. We note that for typical $V_2$ masses in the 
range of a TeV or thereabouts, even the heavy top quarks will be highly 
boosted, and hence the traditional tag of an isolated lepton coming from 
their semileptonic decays will become inefficient \cite{March2006}. The 
high boost can be easily understood since the $p_T$ of the top quarks 
will show a `Jacobian peak' around half of the mass of the heavy 
resonance, subject to minor smearing effects due to the overall boost in 
the laboratory frame. Thus, we must turn to the hadronic decays of top 
quarks, which will lead to a pair of jets having three {\it sub}-jets 
when studied at higher resolution.

\begin{figure}[htb]
\centerline{ \epsfxsize= 5.5 in \epsfysize= 2.8 in \epsfbox{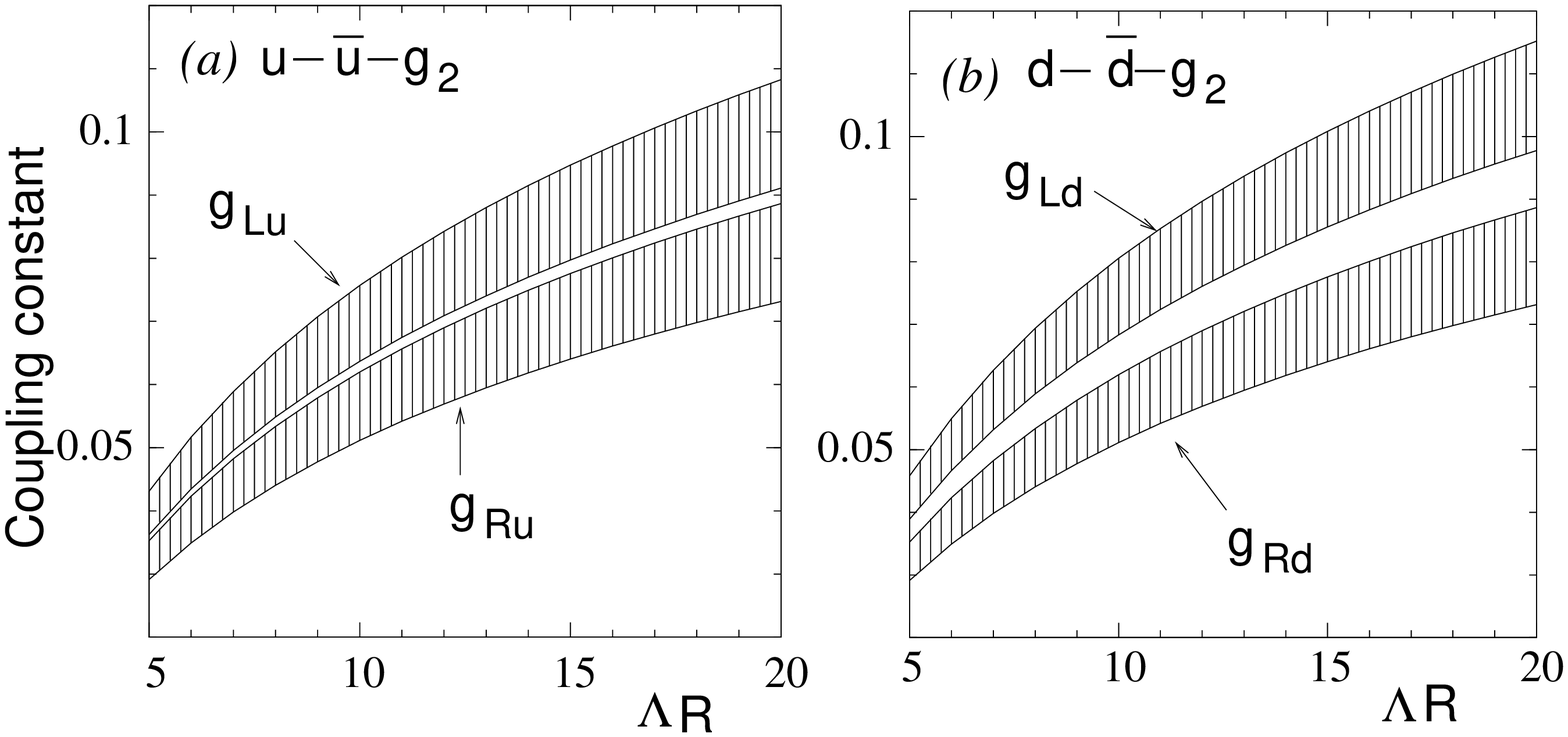} }
\vskip -10pt
\caption{{\footnotesize\it Couplings of the $g_2$ to ($a$) $u$ quarks 
and ($b$) $d$ quarks, as a function of the cutoff parameter $\Lambda R$. 
The width of the curves, indicated by hatching, indicates the weaker 
variation induced by changing $R^{-1}$, which acts through the changed 
running of $\alpha_s$. It is worth noting that the $g_2$ couplings to 
quark pairs are considerably smaller than the corresponding couplings of 
a normal gluon.}}
\label{fig:g2Couplings}
\end{figure}

\bigskip\noindent The rate of production of $V_2$ resonances -- in 
particular, of $g_2$ resonances -- will depend on their couplings to 
light quarks in the proton. Of these, the most important 
$q$--$\bar{q}$--$g_2$ vertices can be parametrised in the form 
\cite{Datta2010}
$$
-i \gamma^\mu 
\left( g_{Lq} \frac{1-\gamma_5}{2} + g_{Rq} \frac{1+\gamma_5}{2} \right)
$$
where $q = u, d$. In the limit where these quarks are massless and the 
masses of all the $n = 1$ fields are degenerate, one can write down 
these form factors $g_{Lq}$ and $g_{Rq}$ in closed form. These 
approximate formulae, given in Ref.~\cite{Datta2010}, are not reproduced 
here in the interests of brevity. However, in 
Figure~\ref{fig:g2Couplings}, we show their variation with the cutoff 
parameter $\Lambda R$. It may be noted that the dependence on the cutoff 
$\Lambda R$ arises from two different sources, viz. the divergent terms 
in the one-loop diagrams, as well as through the running coupling 
constants in the theory, which we evaluate at the resonance scale of $Q 
\approx 2R^{-1}$. This induces, in addition to the logarithmic 
dependence on $\Lambda R$, a weaker logarithmic dependence on the 
compactification scale $R^{-1}$, which is illustrated by the thickness 
of the bands in Figure~\ref{fig:g2Couplings}. These correspond to 
variation of $R^{-1}$ between 400--1400~GeV, the range of interest in 
this work.

\vskip -5pt

\noindent The lesson which is implicit in Figure~2 is the fact that the 
couplings of the $g_2$ to light quarks are rather small, when compared 
with the strong coupling constant $g_s$ which is about 0.25 at these 
energies, and the electromagnetic coupling constant, which is about 0.3. 
This is not really a surprise, though, since we have established that 
the $q$--$\bar{q}$--$g_2$ couplings occur at the one-loop level and 
hence will always be suppressed by a factor of $(16\pi^2)^{-1}$. The 
$g_2$ resonance will be rather small, therefore -- in fact, it will not 
be much larger than a heavy Higgs boson resonance. To distinguish it 
above the background will, then, require a rather fine-grained search, 
as is prescribed in the case of Higgs boson searches..

\begin{figure}[htb]
\centerline{ \epsfxsize= 5.8 in \epsfysize= 3.0 in \epsfbox{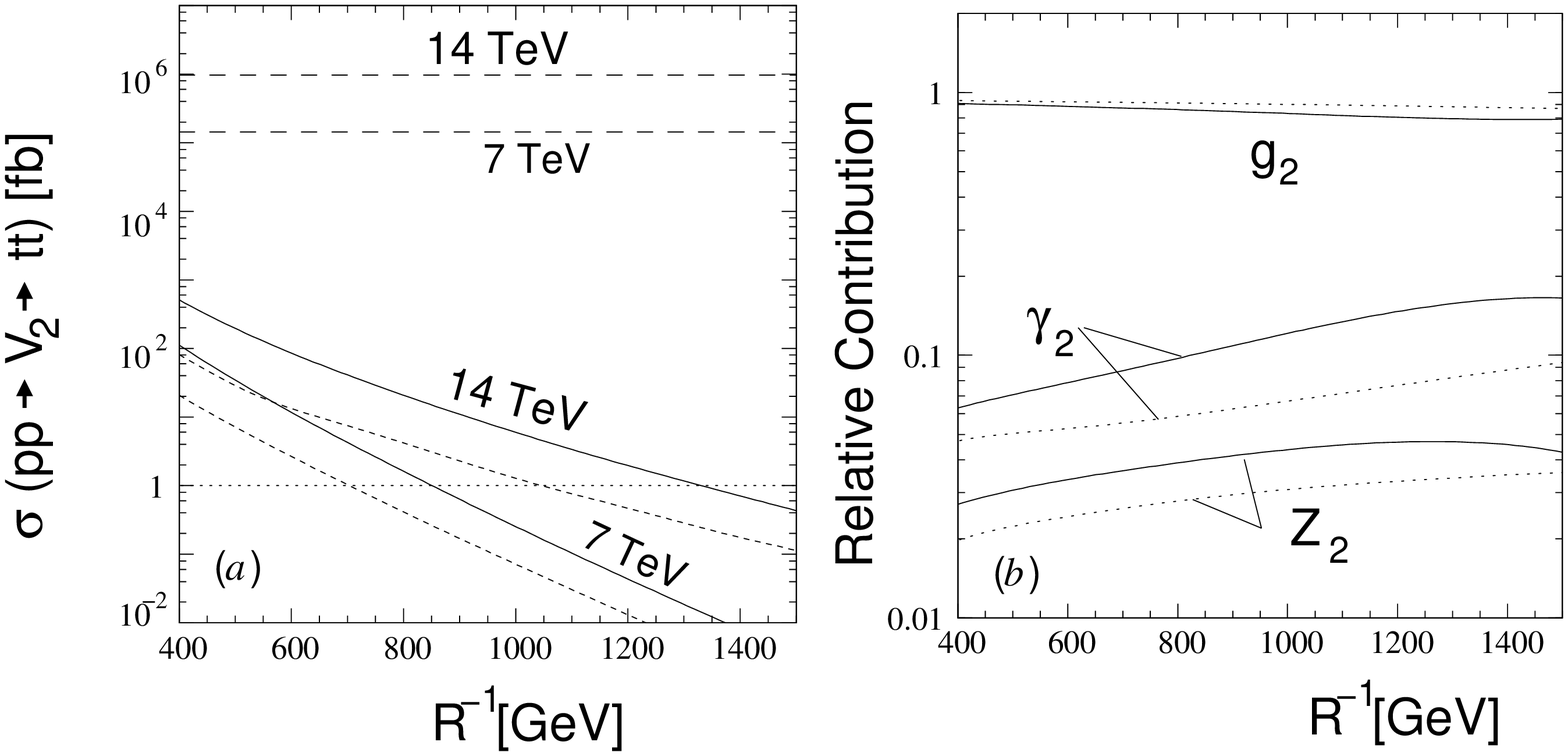} }
\vskip -10pt

\caption{{\footnotesize\it Illustrating {\rm($a$)} the summed resonant 
cross section for $t\bar{t}$ production at the LHC through $V_2 = 
\gamma_2, Z_2$ or $g_2$, as a function of $R^{-1}$. Solid (dotted) lines 
in the lower half correspond to $\Lambda R$ = 20 (5) and the horizontal 
dotted line indicates the projected luminosity reach of the 7~TeV run at 
the LHC \cite{LHCwebsite}. The QCD background is indicated by broken 
lines in the upper half. In {\rm($b$)}, the partial contributions of 
$\gamma_2, Z_2$ and $g_2$ are shown, for $\sqrt{s} = 7$~TeV, where again 
solid (dotted) lines correspond to $\Lambda R$ = 20 (5). The dominance 
of the $g_2$ resonance is obvious.}}
\label{fig:cross section}
\end{figure}

\bigskip\noindent We now take up the case of detecting resonances in the 
$p p \to t \bar{t}$ cross section. We have already mentioned that top 
quark pairs can be produced in large numbers through the resonant 
processes
\begin{eqnarray}
u + \bar{u} \to \gamma_2, \ Z_2^0, \ g_2 \to t + \bar{t} \nonumber \\
d + \bar{d} \to \gamma_2, \ Z_2^0, \ g_2 \to t + \bar{t} \nonumber 
\end{eqnarray}
where $u$ includes the (small) contribution of $c$ quarks in the proton, 
and $d$ includes the (likewise small) contributions of $s$ and $b$ 
quarks in the proton. Even though all these processes arise at the same 
order in perturbation theory, the dominant contribution is still 
expected to be from the intermediate $g_2$ state, which is, after all, 
strongly-interacting. The cross section for $t\bar{t}$ production at the 
LHC through such processes is shown in Figure~\ref{fig:cross section}. 
\\

\vskip -5pt

\noindent On the left side of Figure~\ref{fig:cross section}, we show 
the variation in the cross section for resonant $t\bar{t}$ production as 
a function of the size parameter $R^{-1}$ of the UED-5 model. It may be 
noted that this is the {\it total} cross section sans kinematic cuts and 
efficiency factors, calculated using the software {\sc CalcHEP} 
\cite{CalcHEP}. Solid (dotted) lines correspond to the choices $\Lambda 
R = 20$~(5), forming a band within which we expect the intermediate 
values of $\Lambda R$ to lie. There are two such bands, one for 
$\sqrt{s} = 7$~TeV and one for $\sqrt{s} = 14$~TeV, as indicated on the 
figure itself. It is somehow gratifying that the larger signal 
corresponds to the more popular choice $\Lambda R = 20$. A horizontal 
dotted line at 1~fb indicates the luminosity reach of the 7~TeV run at 
the LHC \cite{LHCwebsite}. It is clear that this run will not be able to 
probe values of $R^{-1}$ much above 600~GeV, and hence, though providing 
a stronger limit than the present Tevatron limit, will hardly improve on 
the low energy bound from radiative $B$ decay. However, in the 14~TeV 
run, assuming an integrated luminosity of 100~fb$^{-1}$, the graph 
indicates that we should be able to probe the UED-5 model all the way to 
$R^{-1} = 1.5$~TeV. Near the top of this plot, two horizontal broken 
lines indicate the NLO SM background, which is enormous, viz. 
82.9~(144)~pb at LO(NLO) for 7~TeV and 406~(968)~pb at LO(NLO) for 
14~TeV, using MSTW-08 structure functions \cite{MSTW}. This background 
will be diminished somewhat in the case of a UED-5, since the running of 
the strong coupling constant $\alpha_s$ is faster in a UED-5 than in the 
SM\footnote{More details are given in the context of 
Figure~\ref{fig:resonance}.}. The nevertheless huge size of this 
background should not deter us from proceeding with this analysis, since 
the resonant cross section will be concentrated in a small bin of 
$t\bar{t}$ invariant mass, which is expected to lie at a value where the 
continuum QCD background falls off to small values. The viability of 
such signals will be demonstrated shortly, but this serves to drive home 
the point that kinematic reconstruction of the final state invariant 
mass plays a pivotal role in this analysis, and that this is feasible 
for highly-boosted top quarks only in the purely hadronic channels.

\bigskip\noindent On the right side, marked ($b$), of 
Figure~\ref{fig:cross section}, we plot the relative contribution to the 
resonant $t\bar{t}$ cross section from intermediate $\gamma_2$, $Z_2$ 
and $g_2$ states, again, as a function of $R^{-1}$. These are drawn 
assuming that $\sqrt{s} = 7$~TeV; for $\sqrt{s} = 14$~TeV, the 
qualitative features will remain the same, but there will be some 
numerical changes due to unequal variation in the corresponding one-loop 
couplings. However, these are not so important. What is significant is 
the fact that the $g_2$ contribution dominates the resonant part of the 
cross section. This is clearly due to the fact that the $g_2$ cross 
section, though occurring at one loop like the others, still has a 
factor of $\alpha_s$ where the $\gamma_2$ and $Z_2$ have a factor of 
$\alpha$. Even at LHC energies, this difference is substantial, and if 
we add to it the various colour factors, the dominance of the $g_2$ 
contribution is quite understandable. Of course, for the same set of 
reasons, the $g_2$ resonance will not be as sharp as the others, but 
this will not matter at the LHC, where the invariant mass binning would 
be crude enough to render all three peaks indistinguishable. The $g_2$ 
dominance encourages us to look for the resonance in purely hadronic 
channels, complementing earlier work \cite{Datta2005} in the dileptonic 
channels.

\bigskip\noindent In order to study resonances in the $t\bar{t}$ cross 
section, therefore, it is first necessary to reconstruct the kinematics 
of top quarks at the LHC. A top quark (or antiquark) decays into a real 
$b$ quark and a real $W$ boson, which has a hadronic branching ratio of 
67.6\%. Thus, the dominant decay of the top quark will be to three jets, 
of which one is a $b$-jet. If the mother top quark is at rest, or is 
moving at low speeds, these three jets will come out widely separated in 
direction. At low energies, e.g. at the Tevatron, the principal tag for 
top quarks is, not the hadronic decay mode, but the rarer cases when the 
$W$ decays leptonically, and a hard isolated lepton can be used to 
trigger the top quark decays within a mass of similar-looking events. 
However, if the top quark is highly boosted --- as is likely to happen 
at the LHC (e.g. a 1~TeV top quark will have a boost parameter $\beta$ 
of about 0.98) --- then all its decay products would tend to be 
collinear and the lepton will appear to be part of a jet, rather than a 
hard decay product in its own right. It has been suggested, therefore, 
that one should turn to the dominant hadronic channels for 
identification of decaying top quarks. Of course, the three jets arising 
from a top quark decay will also tend to be collinear and hence are 
likely to be close enough for the usual jet identification algorithms to 
clump them into a single jet. A smaller angular resolution, however, 
will reveal, in most cases, three sub-jets within the single jet, and 
this kind of sub-structure can be used quite efficiently, to tag top 
quark jets.

\begin{figure}[htb]
\centerline{ \epsfxsize= 7.0in \epsfysize= 2.2in \epsfbox{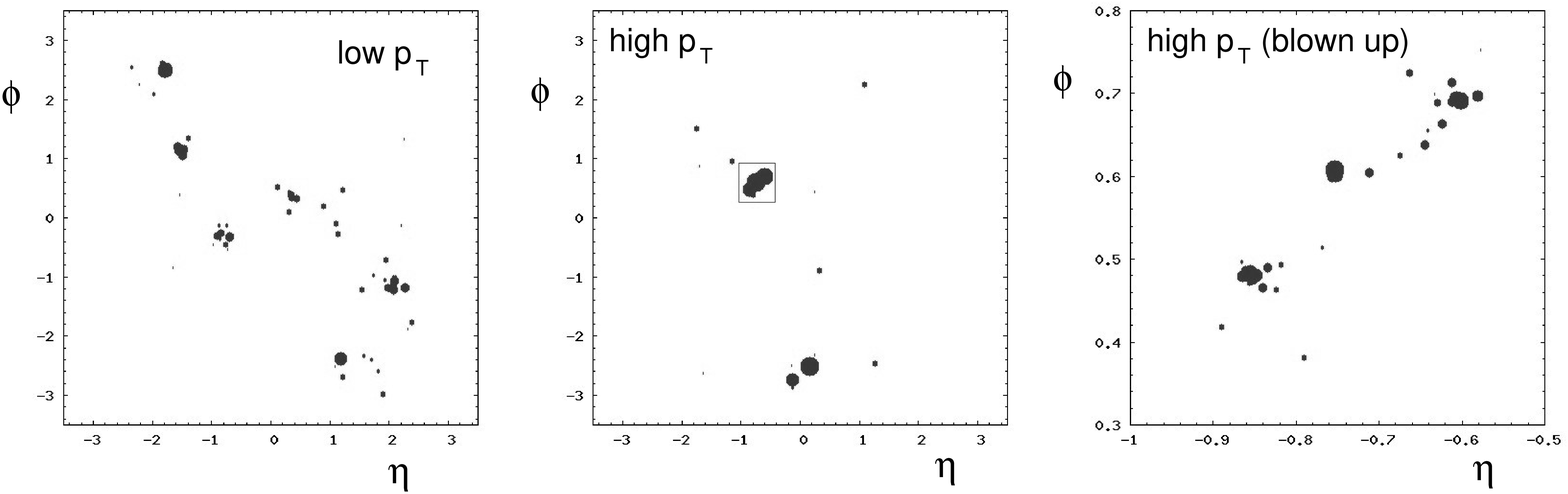} }
\vskip -10pt
\caption{{\footnotesize\it Illustrating the kinematics in the 
$\phi$--$\eta$ plane of hadronic final states in $t\bar{t}$ decay for 
$t$ and $\bar{t}$ with `low $p_T$' $\sim 100$~GeV (extreme left), $t$ 
and $\bar{t}$ with `high $p_T$' $\sim 1$~TeV (centre), and `high $p_T$ 
(blown up)' (extreme right) i.e. the region in the little box enclosing 
the cluster (central box) plotted with high angular resolution. The size 
of the plotted points is indicative of the energy deposit expected in 
the calorimeter.}}
\label{fig:toptagging}
\end{figure}
\vskip -5pt

\noindent In Figure~\ref{fig:toptagging} we show a scatter plot in the 
plane of pseudorapidity $\eta$ and azimuthal angle $\phi$ representing 
some typical hadronic decays of a $t\bar{t}$ final state.  The states in 
question have been generated using the well-known Monte Carlo event 
generator {\sc Pythia}, which simulates the decay and subsequent 
hadronisation of the $t$ and $\bar{t}$ partons. In the box on the left, 
marked `low $p_T$', each little blob indicates a final state hadron 
(mostly pions) arising as the end product of the $t\bar{t}$ decay. The 
plot ranges over the full acceptance range of the LHC hadron 
calorimeter, viz. $-3.5 \leq \eta \leq +3.5$ and $0 \leq \phi < 2\pi$. 
The size of the blob is an indicator of the energy of the 
hadron\footnote{The radius of the blob varies logarithmically with the 
energy, on an arbitrarily-chosen scale. This is a fair simulation of a 
typical calorimeter response to a hadronic shower.} and hence, the tiny 
dots correspond mostly to soft pions etc.. In this plot, one can 
identify six jets simply by inspection and we may expect any reasonable 
clustering algorithm to do the same. Of course, this alone does not 
identify a $t\bar{t}$ event. The full identification will require the 
following further steps:
\begin{enumerate}

\item Two of the jets should be tagged as $b$-jets. If not, the event is 
rejected.

\item Of the remaining four jets, labelled (say) 1, 2, 3 and 4, one 
pairing, i.e. (12)(34) or (13)(24) or (14)(23) should have both pair 
invariant masses in the neighbourhood of the $W$-boson mass, i.e. 
between 70 -- 90~GeV. If not, the event is rejected. We can now 
reconstruct the kinematics of the two $W$ bosons arising in $t\bar{t}$ 
decay.

\item Now, of the two $W$'s, say, $W_1$ and $W_2$ and the two $b$-jets, 
say, $b_1$ and $b_2$, one pairing ($W_1 b_1$)($W_2 b_2$) or ($W_1 
b_2$)($W_2 b_1$) should have both pair invariant masses in the 
neighbourhood of the $t$-quark mass, i.e. between 165 -- 190~GeV. If 
not, the event is rejected.  We can now reconstruct the kinematics of 
the $t$ and the $\bar{t}$.

\item Finally, we reconstruct the invariant mass of the $t\bar{t}$ 
system. This can either be done using the previous results, or simply by 
adding up all the final state momenta for the selected events and 
squaring -- this has the advantage of including all the soft pions etc,, 
which may have been left out of a clustering algorithm. We then check if 
the invariant mass of the entire six-jet system has a peak at some high 
value or not.

\end{enumerate}

\noindent The situation changes if the $t$ and the $\bar{t}$ are highly 
boosted. This is illustrated in the central box, marked `high $p_T$' in 
Figure~\ref{fig:toptagging}. As before, final state hadrons in a {\sc 
Pythia} simulation have been indicated by dark blobs of size varying 
according to the expected energy deposit. In this case, apart from a few 
scattered soft hadronic objects, we see that the main hadronic event 
consists of just two hard transverse jets (note the low value of $\eta$ 
for both clusters), and this is also what a simple-minded jet clustering 
algorithm will tell us. We thus have a pair of $t$ jets, which would 
normally be lost against the enormous QCD dijet background. However -- 
and this is where `top-tagging' techniques come into play -- if we make 
a much more high resolution plot of one of the jet clusters, as shown in 
the box on the right, marked `high $p_T$ (blown up)', then the story 
changes again. The box on the extreme right is a high-resolution 
magnification of the tiny box enclosing the cluster near the centre of 
the middle box (marked `high $p_T$'). Once again, if we neglect soft 
hadrons, this shows a sub-structure with three sub-jets clearly 
identifiable by inspection alone. Thus, if we can pass the putative $t$ 
jets through a suitably-tuned sub-clustering algorithm, which identifies 
three sub-jets as illustrated, and then subject the entire event to a 
filtering process similar to the one described above, we should be able 
to identify $t\bar{t}$ events arising from a resonance, even when the 
$t$ and the $\bar{t}$ are heavily boosted.

\bigskip\noindent Having established the rationale for a top-tagging 
study of dijet events, we now describe the exact process adopted in 
order to implement this. Cross sections for the generation of $V_2$ 
resonances and their decay into $t\bar{t}$ pairs has been done using a 
{\sc CalcHEP}-based program \cite{Datta2010}. The events from these have 
been interfaced with {\sc Pythia} \cite{Pythia} to produce a bunch of 
hadronic final states, similar to those shown in 
Figure~\ref{fig:toptagging}. These are then analysed using the program 
{\sc FastJet} \cite{FastJet}, which is also freely available on the 
Internet. The working of the {\sc CalcHEP} and {\sc Pythia} part of our 
analysis follows standard procedures and is hardly worth discussing 
here. However, as {\sc FastJet} is a relatively new entrant to the field 
of collider studies, a brief review of its working seems appropriate. 
Here we closely follow the discussion of Ref.~\cite{Kaplan2008}, 
mentioning our specific choices of parameters as and when the occasion 
arises.
 
\bigskip\noindent The first step in the algorithm used by the {\sc 
FastJet} program is to cluster the hadronic final states into `jets' of 
angular width $R$ (= 0.8 in our analysis). This is done by using the 
Cambridge-Aachen (CA) algorithm\footnote{This uses the simplest of three 
possible measures of the angular distance between four-vectors, the 
other two being known as the `$k_T$-algorithm' and the `anti-$k_T$ 
algorithm'. Though these other measures are more sophisticated than the 
simpleminded CA measure, a detailed study undertaken by the CMS 
Collaboration \cite{CMS} shows that in practice. all three measures lead 
to similar results.}, i.e. to start with all the four-vectors 
corresponding to the hadronic final states, and then merge the pair 
having the smallest value of $\Delta R \equiv \sqrt{\Delta\eta^2 + 
\Delta\phi^2}$ into a single four vector. The process is repeated until 
there are no four-vectors with $\Delta R < R$. At this state, all the 
surviving four-vectors may be identified as jets. We thus identify a 
multi-jet event. Of course, all information on the original set of 
four-vectors is stored.

\bigskip\noindent The next step is to consider one `jet' at a time and 
to 'decluster' it, using the following algorithm. The clustering process 
mentioned above is reversed, starting from the last two four-vectors to 
be merged. If the final transverse momentum $p_T^{(J)}$ is decomposed as 
$p_T^{(J)} = p_T + p'_T$ we calculate the fractions $p_T/p_T^{(J)}$ and 
$p'_T/p_T^{(J)}$. If one of these ratios comes out to be less than a 
previously-chosen minimum value $\delta_p$ ($= 0.05$ in our analysis), 
the corresponding four-vector is discarded as not being identifiable as 
a sub-jet. The other four-vector, which must then have a large $p_T$ 
ratio, is then subjected to the same declustering process, i.e. it is 
split into the two four-vectors from which it was created by the 
original clustering process. The process then iterates. In every case, 
the ratio is created w.r.t. $p_T^{(J)}$, the total transverse momentum 
of the jet.

\bigskip\noindent The declustering procedure is terminated if one of the 
following situations is encountered:

\begin{enumerate}

\item Both the declustered four-vectors have $p_T/p_T^{(J)} > \delta_p$ 
($= 0.05$ in our analysis).

\item Both the declustered four-vectors have $p_T/p_T^{(J)} < \delta_p$ 
($= 0.05$ in our analysis).

\item The objects are too close, i.e. $|\delta\eta|+|\delta\phi| < 
\delta_r$ ($= 0.1$ in our analysis).

\item There is only one `calorimeter cell' left. For the LHC, it is 
convenient to take a calorimeter cell of 0.1 in both $\eta$ and $\phi$. 
Thus, this condition usually corresponds closely to the previous one.

\end{enumerate}  
If the procedure terminates because of condition 1 above, we consider 
the jet to have two sub-jets. If the procedure terminates due to any of 
the conditions 2, 3 or 4, we consider the original jet to be 
irreducible, i.e. having no sub-jet structure.

\bigskip\noindent In case the original jet is found to have two 
sub-jets, each sub-jet is, in turn treated to the same declustering 
procedure to check for further sub-jets, and the process iterates. The 
procedure terminates when all the sub-jets are found to be irreducible 
by the criteria described above. If we are tagging the original jet for 
a $t$-quark origin, then we select only such cases where there are {\it 
three} distinct sub-jets\footnote{A fourth, less energetic jet is also 
admissible to take care of final state gluon radiation.}. As $b$-tagging 
is not efficient for high-$p_T$ jets, we require only that ($i$) one 
pair of sub-jets should have invariant mass peaking near\footnote{These 
invariant mass criteria are looser for a sub-jet analysis than in the 
case of six distinct jets because the errors are always larger for 
invariant masses of objects which are separated by small angles} the 
$W$-pole, i.e. between 65 -- 95~GeV, and ($ii$) the invariant mass of 
all three sub-jets should peak near the $t$-quark pole, i.e. between 145 
-- 205~GeV. Once an event clears these criteria, we define a `$W$ 
helicity angle' as the angle $\theta_h$, in the rest frame of the 
reconstructed $W$ boson, between the 3-momentum vector corresponding to 
one of the $W$-decay sub-jets and the vector corresponding to the 
overall $t$ (or $\bar{t}$) jet's 3-momentum. To qualify as a $t$ (or 
$\bar{t}$) jet, the helicity angle should satisfy $\cos \theta_h < 0.7$. 
This last criterion is very helpful in preventing QCD jets from being 
mistagged as $t$ jets, the reason being that for such QCD jets, the 
distribution in $\cos \theta_h$ diverges as $(1 - \cos \theta_h)^{-1}$, 
and hence is strongly peaked around $\cos \theta_h \approx 1$. On the 
other hand, for $t$-quark jets, the distribution in $\cos \theta_h$ is 
essentially flat, and hence the signal will be affected only marginally 
by the cut $\cos \theta_h < 0.7$. A sketch of the distributions in $\cos 
\theta_h$ for a $t$-quark jet, a quark jet and a gluon jet may be found 
in Ref.~\cite{Kaplan2008}.

\begin{figure}[htb]
\centerline{ \epsfxsize= 3.2in \epsfysize= 3.0in \epsfbox{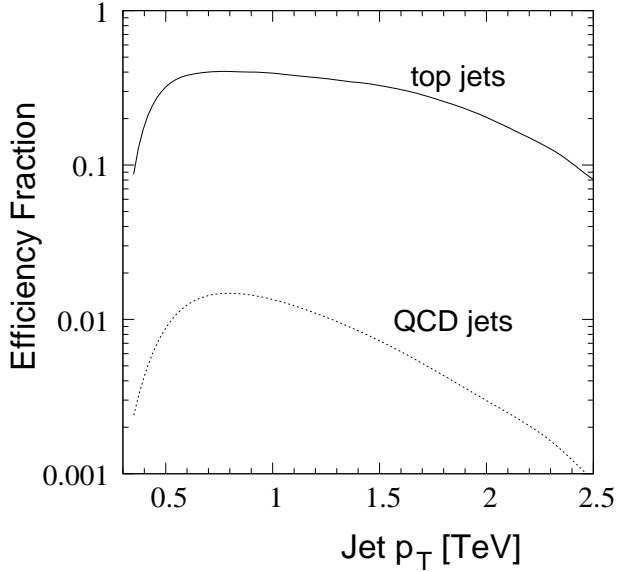} }
\vskip -10pt
\caption{{\footnotesize\it Illustrating the efficiency fractions 
obtained in our implementation of the {\sc FastJet} algorithm. The solid 
line indicates the fraction of $t$-quark jets which are identified as 
such, while the broken line indicates the fraction of light quark or 
gluon jets which are mistagged as $t$-jets.}}
\label{fig:efficiency}
\end{figure}
\vskip -5pt 
  
\noindent In Figure~\ref{fig:efficiency} we plot (solid line) the 
efficiency fraction obtained by us, as a function of the transverse 
momentum $p_T$, for identification of a $t$($\bar{t}$) jet out of a 
genuine sample of $t$($\bar{t}$) jets, together with a plot (broken 
line) of the fraction of QCD dijets that would be mistagged as 
$t$($\bar{t}$) jets. The top-tagging efficiency, shown by the solid 
line, clearly peaks at around 30--40\% for $p_T$ in the range 
0.6--1.5~TeV, after which it falls faster, but still remains more than 
10\%. It is easy to understand the low efficiency for low values of 
$p_T$, for in that case, the $t$-quark will mostly decay into isolated 
jets, and our initial selection of only dijet events will exclude most 
of the genuine $t$-quark decays. Moreover, some of the $t$-jets will be 
fat jets, wider than our acceptance criterion of $R < 0.8$, in which 
case the jet momentum will not carry all of the information about the 
momentum of the parent $t$-quark momentum, resulting in rejection by the 
invariant mass criteria described above. Again, when the $p_T$ is very 
high, i.e. in the neighborhood of 1~TeV, the efficiency falls again, 
because the highly-boosted decay products are so sharply collimated that 
the sub-jets merge and lose their individual identity to an extent that 
the declustering algorithm fails. For the mistagging fraction, shown by 
the broken line which rarely exceeds 1\%, the same arguments can be 
applied. Low-$p_T$ QCD jets will tend to spread out and form either 
multiple jets or fat jets, and hence the number mistaken for $t$-quark 
jets will be less. At high-$p_T$, again, what substructure does arise 
from random fluctuations, will be lost in the sharp collimation of all 
hadronic tracks. It is, in a way, advantageous, that the tagging 
efficiency and the mistagging probability curves have similar shapes. 
For this means that when the signal is low, the number of mistagged 
events will also be low, and when the latter is larger, the signal is 
more healthy.
 
\bigskip\noindent It is important to note that our choice of fixed 
tolerance parameters $\delta_p$ and $\delta_r$ is somewhat looser than 
that those chosen by Kaplan {\it et al} \cite{Kaplan2008} in their 
pioneering discussion of top-tagging techniques. As a result, we have a 
somewhat larger acceptance for $t$ ($\bar{t}$) jets with substructure at 
the cost of a greater acceptance for QCD dijets as well. However, the 
actual differences are small, and when combined with the invariant mass 
and other criteria, lead to very similar results. We feel, therefore, 
that it is reasonable to continue the analysis with fixed tolerance 
parameters.

\bigskip\noindent The actual event selection will be as follows: we 
consider all generated events having two hard jets of $p_T > 500$~GeV 
and no other identifiable activity. If both hard jets can be tagged, 
using the procedure described above, as having a $t$($\bar{t}$) origin 
(this will include a substantial number of mistagged QCD jets, since the 
QCD dijet cross section is very large)), we then construct the invariant 
mass $M(t\bar{t})$ of this pair, expecting $V_2$ resonances to show up 
as bumps in the invariant mass distribution. A plot of the expected 
distribution is shown in Figure~\ref{fig:resonance}, for three different 
values of the size parameter $R^{-1}$. For this graph, and indeed, for 
the rest of this paper, we set $\Lambda R = 20$ and consider only the 
14~TeV run of the LHC.

\bigskip\noindent The shaded histograms in Figure~\ref{fig:resonance}, 
correspond to values of $R^{-1} =$ ($a$)~500~GeV, ($b$)~800~GeV and 
($c$)~1 TeV respectively. Of the un-shaded histograms, the solid line 
corresponds to the Gaussian $1\sigma$ fluctuation in the SM background, 
which has been calculated by adding the $t\bar{t}$ contribution to the 
mistagged contribution from QCD dijets\footnote{These two contributions 
are quite comparable in magnitude, and hence it is essential to consider 
both together.}. It must be noted that once the $R^{-1}$ threshold is 
crossed, the QCD beta function will receive contributions from $n = 1$ 
KK excitations and hence, the running of the coupling constant 
$\alpha_s$ will be different from what we would get in the SM alone. The 
broken line shows what would have been obtained for the fluctuation in 
the background if this effect had not been taken into consideration. 
Obviously, deviations in the two histograms will arise only for 
$M(t\bar{t}) > R^{-1}$.

\begin{figure}[htb]
\centerline{ \epsfxsize= 6.5in \epsfysize= 2.8in \epsfbox{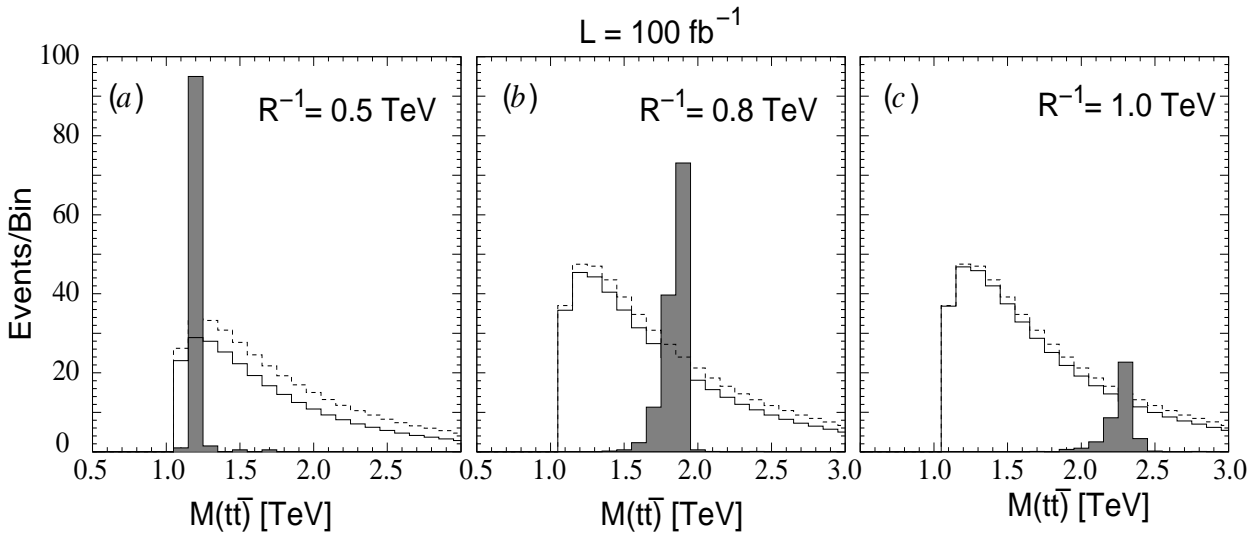} }
\vskip -10pt
\caption{{\footnotesize\it Invariant mass distribution for reconstructed 
$t\bar{t}$ pairs at the 14 TeV run of the LHC, assuming an integrated 
luminosity of 100~fb$^{-1}$, for $R^{-1} =$ ($a$) 500~GeV, ($b$) 800~GeV 
and ($c$) 1 TeV. The shaded histogram represents the signal ($S$) and 
the solid un-shaded histogram represents the Gaussian fluctuation in the 
SM background (continuum $t\bar{t}$ pairs, as well as mistagged dijets) 
($\sqrt{B}$). The broken line shows the fluctuation in the SM background 
if the (faster) UED-5 running of the strong coupling constant $\alpha_s$ 
is ignored.}}
\label{fig:resonance}
\end{figure}
\noindent The background histograms in Figure~\ref{fig:resonance} are 
cut off on the left at $M(t\bar{t}) < 1$~TeV because of the cut $p_T > 
500$~GeV imposed on the triggered jets. The background is largest where 
the identification efficiency and mistagging probabilities are largest, 
i.e. for $M(t\bar{t})$ between 1.0 and 1.5~TeV, and exhibits a monotonic 
decrease as $M(t\bar{t})$ grows larger. This decrease is caused by a 
combination of $s$-channel suppression, falling PDFs, and decreased 
efficiency for very high $p_T$ jets. It allows us to discern a 
diminished signal even for larger values of $R^{-1}$, such as 1~TeV, 
where the $g_2$ resonance lies well over 2~TeV. In order to calculate 
the signal significance, we calculate a $\chi^2$ for the deviation from 
the continuum UED-5 prediction. This is given by the formula
\begin{equation}
\chi^2 = \sum_{\langle i\rangle} \frac{S_i}{\sqrt{B_i}}
\end{equation}
where $S_i$ and $B_i$ refer to numbers of events in the $i$-th bin from 
the resonant signal ($S$) and the continuum background ($B$) 
respectively, and the sum $\langle i\rangle$ runs only over bins where 
$S_i/\sqrt{B_i} \geq 3$. This value of $\chi^2$ is then compared with 
the corresponding number predicted for Gaussian random fluctuations in 
the same number of bins at a given confidence level (CL). Taking this as 
a general procedure, and noting that $\chi^2 \propto \sqrt{L}$, where 
$L$ is the integrated luminosity, we can obtain significance figures in 
terms of CL for the entire range of $R^{-1}$ values, for any luminosity 
estimate.

\begin{figure}[htb]
\centerline{ \epsfxsize= 3.5in \epsfysize= 3.2in \epsfbox{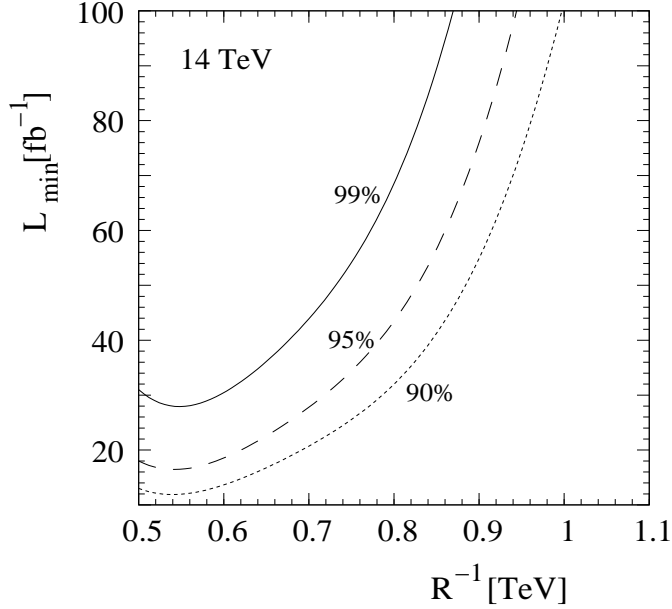} }
\vskip -10pt
\caption{{\footnotesize\it Showing the minimum integrated luminosity 
$L_{min}$ required at the 14~TeV LHC to obtain a $g_2$ resonance peak in 
$t\bar{t}$ signals at the confidence level (CL) of 90\% (dotted line), 
95\% (dashes) and 99\% (solid line).}}
\label{fig:luminosity}
\end{figure}
\vskip -5pt

\noindent In Figure~\ref{fig:luminosity}, we show, using the above 
procedure, the luminosity reach of the LHC, running at $\sqrt{s} = 
14$~TeV, in order to get a significance of 90\% (dotted line), 95\% 
(dashes) and 99\% (solid line). It is clear that even with the 
not-unreasonable estimate of 50~fb$^{-1}$ for the luminosity, we could 
obtain at least a 95\% CL signal all the way up to about $R^{-1} = 
850$~GeV, which is close to the upper bound derived from the dark matter 
constraint. If $R^{-1}$ happens to be smaller, in the neighbourhood of 
600~GeV, then even in the early runs at 14~TeV, we may expect a 
$2\sigma$ resonance peak in the $t\bar{t}$ cross section. Combined with 
missing energy signals (such as a trilepton plus jets plus missing 
energy), this would then be an unambiguous signal for the existence of a 
UED. Values of $R^{-1}$ above a TeV would be accessible only if the much 
talked-about luminosity upgrade \cite{sLHC} does happen, and hence, at 
the moment, may be considered as being only a remote possibility.

\begin{figure}[htb]
\centerline{ \epsfxsize= 3.2in \epsfysize= 3.0in \epsfbox{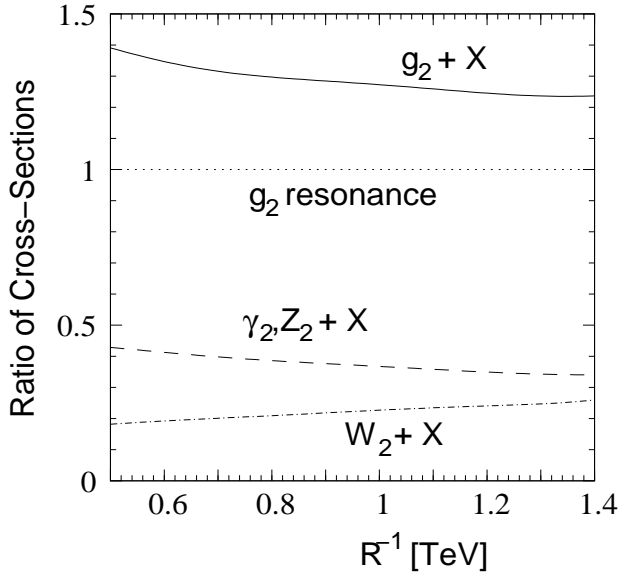} }
\vskip -10pt
\caption{{\footnotesize\it Showing the ratio of $t\bar{t}$ cross 
sections for associated $V_2$ production as a fraction of the resonant 
$G_2$ production cross section shown in Figure~\ref{fig:cross section}. 
The solid line illustrates the sum of both resonant and associated cross 
sections for $g_2$, whereas the others are just the associated cross 
sections.}}
\label{fig:associated}
\end{figure}

\bigskip\noindent In view of the rather promising results presented 
above, it is our belief that the reconstruction of $g_2$ resonances from 
$t\bar{t}$ final states could be the best method of detecting $n = 2$ 
states in a UED-5 theory. The luminosity reach in this channel is fully 
comparable with that predicted in the dilepton channel by Datta {\it el 
al} \cite{Datta2005}, and may even be marginally better. We now turn to 
the issue of {\it associated} production of $V_2$ resonances at the LHC. 
This can arise as a result of several processes, such as $p + p \to q + 
V_2$, $p + p \to g + V_2$, and $p + p \to q + q_2$ where $V_2$ is any 
one of the vector bosons $\gamma_2, Z_2, W_2^\pm$ or $g_2$ and $q_2$ may 
be an SU(2) singlet or doublet quark. In turn, the heavier $g_2$ or 
$q_2$ may decay through cascades to the lighter $\gamma_2, Z_2$ and 
$W_2^\pm$.  We may denote these associated production processes, in 
general, as $p + p \to V_2 + X$, where $X$ is inclusive of leptons, jets 
and missing energy. Cross sections for these can be readily calculated 
using the {\sc CalcHEP} routines mentioned before. In 
Figure~\ref{fig:associated} we plot the cross sections, at the 14~TeV 
LHC, as a ratio with the resonant $g_2$ production cross ection, taking 
only the decay of $\gamma_2, Z_2$ and $g_2$ to $t\bar{t}$ pairs and the 
decay of $W_2$ to $t\bar{b}$ or $\bar{t}b$ pairs (i.e. the production 
cross section of $p + p \to V_2 + X$ are convoluted with the branching 
ratios of $\gamma_2, Z_2, g_2 \to t + \bar{t}$ or $W_2 \to t + \bar{b}$ 
as the case may be).

\vskip -5pt

\noindent The horizontal dotted line in Figure~\ref{fig:associated} 
represents the resonant $g_2$ cross section (for $\Lambda R = 20$) for 
$p + p \to t + \bar{t}$ as a ratio with itself, i.e. unity. The solid 
line represents its ratio with the sum of the associated and resonant 
production cross sections, i.e. we plot
$$
\frac{ \sigma(pp \to g_2X) + \sigma(pp \to g_2)}{\sigma(pp \to g_2)}
$$ 
as a function of the mass parameter $R^{-1}$. It may be seen that an 
enhancement in the $g_2$ signal varying from a factor of about a half to 
a quarter may be achieved by adding on the associated production. 
However, this turns out not to be the best policy, for the following 
reason. If the signal in question is $p + p \to t + \bar{t} + X$, then 
we must also consider the background for this. There are a host of 
processes which can contribute to this, including QCD processes such as 
three-jet production, with mistagging as $t$-jets. These are likely to 
increase the background by a considerably larger factor than the 
enhancement factor we may get for a signal, and would lead, therefore, 
to a reduced significance and lower discovery limits than what we have 
already obtained. We advocate, therefore, that the search should 
concentrate on the clean signal when there are just two hard $t$-jets 
and nothing else in the final state, and do not pursue the issue of 
associated production in the $g_2 + X$ channel any further.

\bigskip\noindent The dashes in Figure~\ref{fig:associated} indicate the 
cross section for associated production of the electroweak $\gamma_2$ 
and $Z_2$ in $p + p$ collisions, once again, as a ratio with the $g_2$ 
resonant cross section. These cross sections are quite large, varying 
from about 30 - 45\%, and certainly much larger than the resonant cross 
sections for $\gamma_2$ and $Z_2$, as shown in Figure~\ref{fig:cross 
section}. This phenomenon has already been noted and used skilfully in 
Ref.~\cite{Datta2005} where the $\gamma_2$ and $Z_2$ have been studied 
in the context of their purely leptonic decays. In the present case, it 
is not very meaningful to consider the $t\bar{t}$ decays of the 
$\gamma_2$ and $Z_2$, since these will come associated with several 
possibilities $X$, and would again be swamped by the large background to 
a $t\bar{t}X$ final state. Finally, the dot-dashed line near the bottom 
of Figure~\ref{fig:cross section} shows the cross section for $p + p \to 
W_2^\pm + X$, again, as a ratio with the resonant $g_2$ cross section. 
This cross section is smaller and in any case, the only hadronic channel 
in which we can even think of searching for this resonance is $W_2 \to 
t\bar{b}$ or $\bar{t}b$. However, at these energies, we have seen that 
$b$-tagging is not possible, so that the $b$-jets will be 
indistinguishable from a light quark jet. Accordingly the selection of 
events will depend on the tagging of a single $t$ quark (or antiquark). 
When we compare this with the $t\bar{t}$ case, and consider the 
tagging/mistagging efficiencies, we see that the signal will be 
increased roughly three times due to the use of one efficiency factor 
rather than two, but the background will increase by two orders of 
magnitude since only one mistagging probability will be required instead 
of two. Accordingly, any $W_2$ resonance will be completely lost against 
the mistagged dijet background. It may be more useful to look, 
therefore, for the decay $W_2 \to \ell + \nu$, but this will be hampered 
by its very low branching ratio. Moreover, a recent study 
\cite{Najafabadi} seems to indicate destructive interference between the 
$W_2$ contribution to $\ell \nu$ and its SM counterpart. All in all, 
associated $V_2$ production is not nearly as viable a signal as the 
resonant $g_2$, unless we look in the purely dilepton channel with a 
high luminosity of 100~fb$^{-1}$ or more.

\bigskip\noindent To conclude, therefore, we have studied $t\bar{t}$ 
production through a resonant $g_2$ in the simplest model with a 
universal extra dimension. Using recently-developed techniques of top 
quark tagging for jets with high transverse momentum, we show that it is 
possible to isolate $g_2$ resonances and obtain an observable signal for 
much of the parameter space of the model which is interesting in the 
context of a dark matter candidate. This will be possible at the 14~TeV 
run of the LHC, as the 7~TeV run will not collect enough luminosity for 
the signal to be discernible over the background.  For very large values 
of the size parameter $R^{-1}$, however, it may not be possible to 
identify $g_2$ resonances with the luminosity actually available at the 
LHC. We have also studied associated production of $V_2$ bosons and 
shown that this is not very useful if we are triggering on top quark 
final states. Our work complements and extends the search for $\gamma_2, 
Z_2$ resonances explored in the 2005 work of Datta {\it et al} 
\cite{Datta2005}.  As $t\bar{t}$ final states will be one of the primary 
channels of interest at the LHC, one may hope that even in the early 
days of the 14~TeV run, we would be able to start accessing the 
parameter space of a UED model in a simple and efficient way.

\bigskip\noindent {\it Acknowledgments}: {\footnotesize The authors 
gratefully acknowledge discussions with Anirban Kundu and with Gobinda 
Majumdar of the CMS Collaboration. }

\end{document}